\documentclass[11pt,dvips]{article}

\usepackage{picinpar}
\usepackage{wrapfig}
\usepackage{floatflt}
\usepackage{psfig}
\makeatletter
\renewcommand{\section}{\@startsection{section}{1}{0in}
	{0.4\baselineskip}{0.1\baselineskip}{\Large\bf}}
\renewcommand{\subsection}{\@startsection{subsection}{2}{0in}
	{0.25\baselineskip}{-\baselineskip}{\large\bf}}
\renewcommand{\subsubsection}{\@startsection{subsubsection}{3}{0in}
	{0.1\baselineskip}{-\baselineskip}{\normalsize\bf}}

\makeatother
\newcommand{\vol}[2]{$\,$\rm #1\rm , #2.}     
\def\aap{Astron. Astr.}                         
\def\apj{Astrophys. J.}                         
\def\apjl{Astrophys. J.}                        
\def\jgr{J. Geophys. Res.}                      
\def\ssr{Space Sci. Rev.}                       

\begin{document}

%
\thispagestyle{myheadings}
%
\markright{SH.4.1.01}
\begin{center}
%
{\LARGE \bf The Production of Anomalous Cosmic Rays by the Solar Wind Termination Shock}
\end{center}

\begin{center}
%
%
{\bf Frank C. Jones$^1$, Matthew G. Baring$^{1,2}$ and Donald C. Ellison$^3$\\}
{\it $^{1}$Laboratory for High Energy Astrophysics, NASA/GSFC, Greenbelt, MD 20771, USA\\
$^{2}$Universities Space Research Association\\
$^{3}$Department of Physics, North Carolina State University, Box 8202, Raleigh NC 27695, U.S.A.}
\end{center}

\begin{center}
{\large \bf Abstract\\}
\end{center}
\vspace{-0.5ex}
%
%
We have modeled the injection and  acceleration of pickup ions at the
solar wind termination shock and investigated the parameters needed to
produce the observed Anomalous Cosmic Ray (ACR) fluxes. A non-linear
Monte Carlo technique was employed which in effect solves the Boltzmann
equation and is not restricted to near isotropic particle distribution
functions.  This technique models the injection of thermal and pickup
ions, the acceleration of these ions, and the determination of the
shock structure under the influence of the accelerated ions.  The
essential effects of injection are treated in a mostly self-consistent
manner, including  effects from shock obliquity, cross-field diffusion,
and pitch-angle scattering. Using recent determinations of pickup ion
densities, we are able to match the absolute flux of hydrogen in the
ACRs  by assuming that pickup ion scattering mean free paths, at the
termination shock,  are much less than an AU and that moderately strong
cross-field diffusion occurs. Simultaneously, we match the flux {\it
ratios} He$^{+}$/H$^+$\ or O$^{+}$/H$^+$\ to within a factor $\sim 3$.
If the conditions of strong scattering apply, {\it no
pre-termination-shock injection phase is required} and  the injection
and acceleration of pickup ions at the termination shock is totally
analogous to the injection and acceleration of ions at highly oblique
interplanetary shocks recently observed by the Ulysses spacecraft.
%

\vspace{1ex}

%
%
\section{Introduction}
 \label{sec:intro}
It is believed that Anomalous Cosmic Rays (ACRs) originate as
interstellar pickup ions (Fisk, Kozlovsky, \& Ramaty 1974) which are
accelerated at the solar wind termination shock (Pesses, Jokipii, \&
Eichler 1981).  Such ions originate as neutrals that are swept into the
solar system from the external interstellar medium, and subsequently
ionized by the solar UV flux or by charge exchange with solar wind
ions.  Recent observations of pickup ions by the Ulysses spacecraft
(e.g.  Gloeckler et al.  1993) adds to the indirect evidence for this
scenario, which by now has become quite compelling.  However, one
essential element of the process, namely how pickup ions are first
injected into the acceleration mechanism, has engendered controversy.
We show here that standard and well-tested assumptions of diffusive
shock acceleration allow the direct injection and acceleration of
pickup ions without a pre-injection stage.

We have employed our Monte Carlo simulation code (e.g.  Ellison,
Baring, \& Jones 1996) to study the physical parameters that the solar
wind termination shock must have in order to produce the observed ACR
fluxes.  For input at the termination shock, we use a standard
expression for the shape of the isotropic pickup ion phase-space
distribution based on the derivation of Vasyliunas \& Siscoe  (1976)
(e.g.  Gloeckler et al.  1993, 1994; le Roux, Potgieter, \& Ptuskin
1996), and normalize this to the values reported by Cummings \& Stone
(1996) for the interstellar ion flux in the heliosphere.  For typical
cases, we require that $\lambda_{\parallel} \sim$ 5-10 $r_g$, where
$\lambda_{\parallel}$ is the ratio of the scattering mean free path
parallel to the mean magnetic field and $r_g$ is the ion gyroradius.
This length scale of diffusion parallel to the field seems fairly
typical of that inferred in the vicinity of planetary bow shocks
(Ellison, M\"obius, \& Paschmann  1990), interplanetary shocks (Baring
et al.  1997), supernova shocks (Achterberg, Blandford, \& Reynolds
1994), and that found in hybrid simulations of quasi-parallel shocks
(e.g.  Giacalone, et al.  1993).

Our principal result is that we can model the {\it absolute} hydrogen
flux with no pre-acceleration.  This is in clear contradiction with the
conclusions of most previous work (discussed in Ellison, Jones and
Baring 1999) addressing pickup ion injection at the termination shock.
We are somewhat less successful in matching the ACR flux {\it ratios},
He$^{+}$/H$^{+}$\ and O$^{+}$/H$^{+}$, seeing less enhancement based on
mass/charge than reported by Cummings \& Stone  (1996).  We do,
however, match the ratios to within a factor of $\sim 5$, a relatively
small difference given the uncertainties of extrapolating flux
densities to the termination shock and the possibility that
species-dependent heating or pre-acceleration could occur in the solar
wind before pickup ions reach the termination shock.  Details of the
results presented here can be found in Ellison, Jones and Baring
(1999, hereafter EBJ99).

\section{Results}
 \label{sec:results}
The Monte Carlo technique we use here has been described in detail in
Ellison, Baring, \& Jones (1996).  Briefly, we have developed a
technique for calculating the structure of a plane, steady-state,
collisionless shock of arbitrary obliquity and arbitrary sonic and
Alfv\'en Mach numbers greater than one. We include the injection and
acceleration of ions directly from the background plasma and assume
that, with the exception of pickup ions, no {\it ad hoc} population of
superthermal seed particles is present.  The model assumes that the
background plasma, including accelerated particles, and magnetic fields
are dynamically important and their effects are included in determining
the shock structure, i.e. shock-smoothing interior to the termination
shock.  The three most abundant ACR species, H$^{+}$, He$^{+}$, and
O$^{+}$, are included self-consistently in the determination of the
shock structure.  Since our Monte Carlo  technique has not yet been
generalized to spherical geometry, we are forced to assume that the
termination shock is plane.  However, the most important process we
investigate, the injection of pickup ions, occurs locally and will not
be seriously affected by this approximation.  

\begin{figure}
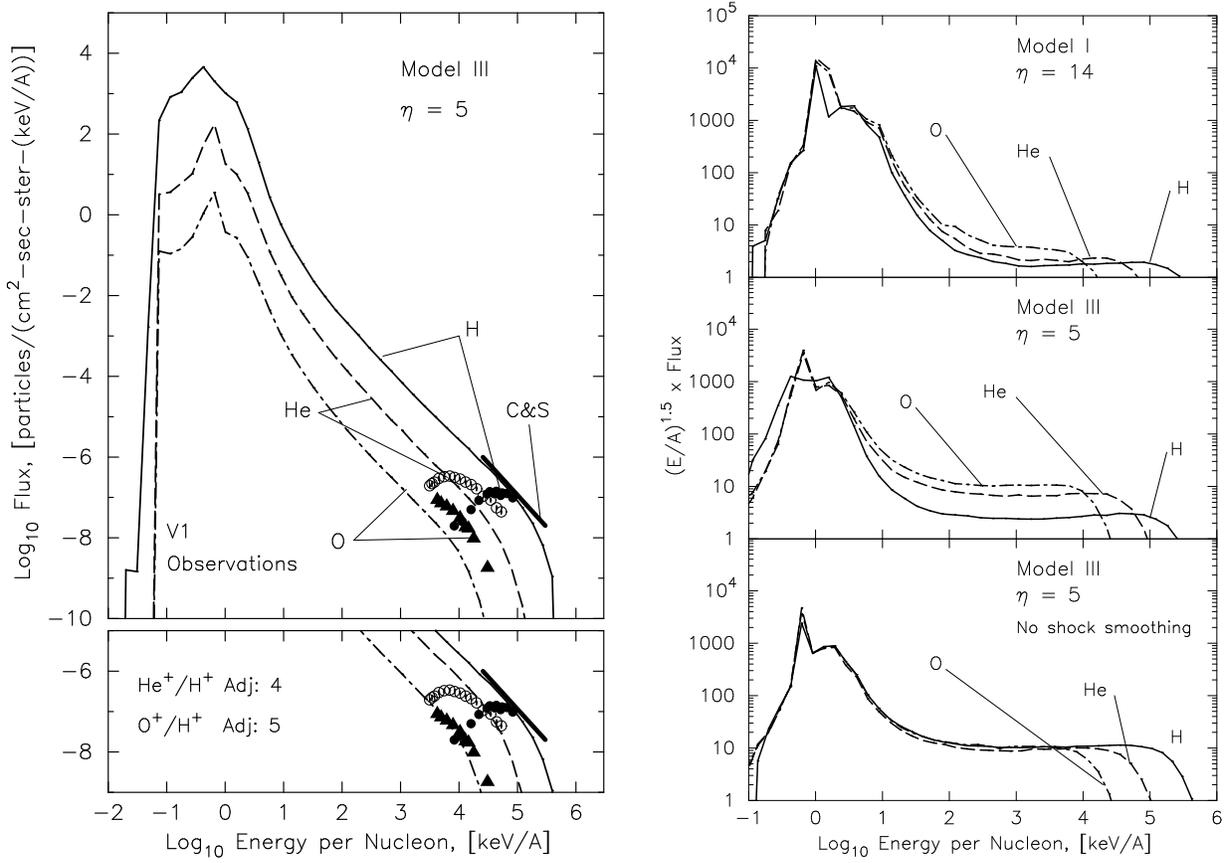

\centerline{\psfig{figure=apj99ejb_ts_f6.ps,width=8.0cm}\hskip 0.7truecm
              \psfig{figure=apj99ejb_ts_f7.ps,width=7.5cm}}
\caption{
{\it (Left Panel)}
Comparison of Voyager 1 observations of ACR H, He, and O (made during
1994/157-313 when V1 was at an average radial location of $\sim 57$ AU)
to Model III spectra (see Ellison, Jones \& Baring 1999 for a
discussion of models) calculated {\it at the termination shock}. The
model spectra have an absolute normalization determined by the
injection parameters, i.e.  $n_{\rm p,TS} V_{\rm sw} = 5.5\times
10^{4}$ cm$^{-2}$ s$^{-1}$ \ for the protons and corresponding values
for the He and O. The value of $\eta =\lambda/r_g$ has been chosen to
give a general fit to the intensities of the observed ACR's.  The sharp
thermal peaks show the relatively cold solar wind ions that have not
yet thermalized. As the observation position is moved downstream, these
peaks broaden.  Note that the H thermal peak intensity is $\sim 11$
orders of magnitude above the observed ACR intensity. The heavy solid
line is the Cummings \& Stone (1996) estimate for the ACR proton
intensity at the termination shock.  In the bottom left panel, we have
individually adjusted the normalizations to match the ACR observations.
The relative adjustments for He$^{+}$\ and O$^{+}$\ are labeled.
{\it (Right Panel)}
Spectra from Models I and III renormalized and multiplied by
$(E/A)^{1.5}$. In each case, we have normalized all spectra to the same
pickup ion density, i.e.  for Model I we have multiplied the He
spectrum by $n_{\rm p}^{\rm pu}/n_{\rm He}^{\rm pu} \simeq 43$ and the
oxygen by $n_{\rm p}^{\rm pu}/n_{\rm O}^{\rm pu} \simeq 1800$, and for
Model III, we have multiplied the He spectrum by $n_{\hbox{\rm
p}}^{\hbox{\rm pu}}/n_{\hbox{\rm He}}^{\hbox{\rm pu}} \simeq 170$ and
the oxygen by $n_{\hbox{\rm p}}^{\hbox{\rm pu}}/n_{\hbox{\rm
O}}^{\hbox{\rm pu}} \simeq 7000$.  In the top two panels, the
self-consistent smooth shock is used to produce the spectra and a clear
$A/Q$ enhancement of He$^{+}$\ or O$^{+}$\ to H$^{+}$\ is seen.  In the
bottom panel, we determined the spectra using the test-particle,
discontinuous shock and essentially no enhancement (other than
statistical variations) is present.
 }
   \label{fig:apj99ejb_figs67}
\end{figure}

Details of the solar wind and pick-up ion distributions and parameters
at the termination shock used in our model are presented in Ellison,
Jones and Baring (1999), as are parameters for three models used to
explore a variety of possible situations.  An example of our results is
depicted in Figure~1.  In the left panel, we depict spectra resulting
from a low Mach number (i.e. weak) shock of obliquity $\Theta_{\rm
Bn1}=87^\circ$ with compression ratio $r\simeq 2.8$, produced using a
slow solar wind speed of 360 km/sec at the termination shock.  Due to
the intrinsically steep spectra produced, we require a moderately low
value of the ratio $\eta =\lambda_{\parallel}/r_g$ of the ion mean free
path to its gyroradius to approximately match the observed ACR data
from the Voyager 1 spacecraft (Cummings \& Stone 1996).  This value is
quite possible given the low $\eta$ inferred from spectral data at
nearby interplanetary shocks (Baring et al. 1997).  Models with
stronger shocks can achieve similar injection efficiencies with
considerably higher $\eta$.  Furthermore, since the flux of
model-generated ACRs is strongly anti-correlated with the shock
obliquity, we find that (EBJ99) a shock of obliquity $\Theta_{\rm
Bn1}=80^\circ$ can roughly replicate the observed flux levels in the
limit of Bohm diffusion ($\eta =1$), using injection in the absence of
pick-up ions, by directly accelerating the solar wind population.

The spectra exhibited in the left panel of Figure~1 possess the
so-called mass ($Am_{\rm p}$) to charge ($Qe$) enhancement.  This
effect, namely that the acceleration efficiency of shocks that are
smoothed by the pressure of accelerated particles is an increasing
function of $A/Q$, has been known for some time (e.g.  Eichler 1979;
Ellison, Jones, \& Eichler 1981) in quasi-parallel scenarios.  It
depends only on the conservation of momentum and a spatial diffusion
coefficient which is an increasing function of energy, and occurs
because non-relativistic ions with larger $A/Q$ (i.e. larger
rigidities) have longer upstream diffusion lengths, at a given energy
per nucleon.  The fact that the shock is smoothed means that the high
$A/Q$ particles `feel' a larger effective compression ratio and are
accelerated more efficiently and, the greater the smoothing, the
greater the enhancement.  Enhancements have been confirmed at the
quasi-parallel Earth bow shock (i.e.  Ellison, M\"obius, \& Paschmann
1990) and should occur regardless of the shock obliquity {\it as long
as the shock is smoothed}.  To demonstrate its appearance in the
simulation results here, we provide a ``normalized'' representation of
spectra from two of our models in the right panel of Figure~1, together
with a simulational example where shock smoothing has been artificially
suppressed and a sharp discontinuity retained.  While there is a clear
$A/Q$ enhancement, it is insufficient to explain the observed
abundances in the Voyager 1 data of Cummings and Stone (1996).

One consistency check with our modeling is to verify that the time we
compute for acceleration is consistent with experimental limits
obtained from charge-stripping rates.  Such ionization is relevant to
species heavier than He (whose stripping timescales are long), in this
case oxygen.  Adams \& Leising (1991) showed that 10 MeV/A singly
charged oxygen ions will be further stripped, in conflict with
observations, if they propagate more than $\sim 0.2$ pc in the local
interstellar medium.  This corresponds to a stripping timescale of
around 3--5 years at 10 MeV/A.  BJE99 demonstrates that the
acceleration timescales for shock obliquities exceeding around
$70^\circ$ fall short of this observational constraint, thereby
validating our assumption that oxygen is singly-ionized at all (but
perhaps the highest; see Mewaldt et al. 1996) energies.

In conclusion, using standard solar wind quantities and basic
microphysical parameters, we have shown that diffusive shock
acceleration operating at the termination shock can account for
observed ACR proton fluxes by directly accelerating pickup ions from
solar wind speeds to $\sim 150$ MeV, without any imposition of a
pre-injection phase.  The only requirements for direct injection is
that local magnetic turbulence exists (presumably self-generated) such
that $\kappa_{\perp}/\kappa_{\parallel} \gg 10^{-3}$ and that
$\lambda_{\parallel}$, for pickup ions injected at the shock, is a
small fraction of an AU.  These criteria are not difficult to satisfy
in heliospheric environments.  From our results, we conclude that the
acceleration process at the termination shock is, as far as limited
observations allow us to determine, identical in all important respects
to diffusive particle acceleration observed at inner heliospheric
systems such as the Earth bow shock and travelling interplanetary
shocks.

%
\vspace{1ex}
\begin{center}
{\Large\bf References}
\end{center}
  Achterberg, A., Blandford, R.~D. \& Reynolds, S.~P. 1994, \aap\vol{281}{220}
\\
  Adams, J.~H. \& Leising, M.~D. 1991, in Proc. 22nd ICRC (Dublin), \vol{3}{304}
\\
  Baring, M.~G., Ogilvie, K.~W., Ellison, D.~C., \& Forsyth, R.~J. 1997,
  \apj\vol{476}{889}
\\
  Cummings, A.~C., \& Stone, E.~C. 1996, \ssr\vol{78}{117}
\\
  Eichler, D. 1979, \apj\vol{229}{419}
\\
  Ellison, D.~C., Baring, M.~G., \& Jones, F.~C. 1996, \apj\vol{473}{1029}
\\
  Ellison, D.~C., Jones, F.~C., \& Baring, M.~G. 1999, \apj\vol{512}{403}
\\
  Ellison, D.~C., Jones, F.~C., \& Eichler, D. 1981,
  Journal of Geophysics, \vol{50}{110}
\\
  Ellison, D.~C., M\"obius, E., \& Paschmann, G. 1990, \apj\vol{352}{376}
\\
  Fisk, L.~A., Kozlovsky, B., \& Ramaty, R. 1974, \apjl\vol{190}{L35}
\\
  Giacalone, J., Burgess, D., Schwartz, S.~J., \& Ellison, D.~C. 1993, 
  \apj\vol{402}{550}
\\
  Gloeckler, G., Geiss, J., Fisk, L.~A., Galvin, A.~B., Ipavich, F.~M.,
  Ogilvie, K.~W., von Steiger, R., \\
 \hbox{\hskip 10pt}  \& Wilken, B. 1993, Science, \vol{261}{70}
\\
  Gloeckler, G., Geiss, J., Roelof, E.~C., Fisk, L.~A., Ipavich, F.~M.,
  Ogilvie, K.~W., Lanzerotti, L.~J., \\
 \hbox{\hskip 10pt}  von Steiger, R., \& Wilken, B. 1994,
  \jgr\vol{99}{17,637}
\\
  Le Roux, J.~A., Potgieter, M.~S., \& Ptuskin, V.~S. 1996, \jgr\vol{101}{4791}
\\
  Mewaldt, R.~A., Selesnick, R.~S., Cummings, J.~R., Stone, E.~C.,
  \& von Rosenvinge, T.~T. 1996, \\
 \hbox{\hskip 10pt}  \apjl\vol{466}{L43}
\\
  Pesses, M.~E., Jokipii, J.~R., \& Eichler, D. 1981, \apjl\vol{246}{L85}
\\
  Vasyliunas, V.~M., \& Siscoe, G.~L. 1976, \jgr\vol{81}{1247}
\\


\end{document}